\documentclass{article}
\usepackage{parskip}
\usepackage{float}
\usepackage{amsmath}

\usepackage[english]{babel}

\usepackage[a4paper,top=2cm,bottom=2cm,left=3cm,right=3cm,marginparwidth=1.75cm]{geometry}

\usepackage{amsmath}
\usepackage{graphicx}
\usepackage[colorlinks=true, allcolors=blue]{hyperref}

\title{How to design a network architecture using availability}
\author{Gilbert Moïsio \\ Network \& Methodology Senior Consultant \\ gmoisio@gmail.com}
\date{March 30, 2013}

\begin{document}
\maketitle

\begin{abstract}
The best way to design a network is to take into account Availability values and Capacity Planning.  You already saw Availability expressed with numbers such as $99.99\% = 0.9999 = 1-10^{-4}$ or Unavailability expressed with $1-(1-10^{-4})=10^{-4}$. The purpose of this document is to introduce the way to compute Availability values using Reliability Block Diagrams.
\end{abstract}

\section{Introduction}

The proper functioning of the infrastructure and its availability are fundamental elements for the smooth running of an establishment.

Instantaneous availability \emph{A(t)} is the ability (probability) of an entity to be able to perform a required function under given conditions, at a given time \emph{t}, assuming that the supply of the necessary external means is ensured. It takes into account both reliability and maintainability.

Maintainability is, under given conditions of use, the ability (probability) of an item to be maintained or returned to service over a given interval of time, in a state in which it can perform a required function, when maintenance is performed under given conditions with prescribed procedures and means.

The definition of availability is a definition of the IEC, it is modeled on the reliability one but the temporal aspect is fundamentally different since we are interested in a state at a given instant \emph{t} and not in a duration. Operation at time \emph{t} does not necessarily require operation on \emph{[0, t]}.

\section{Availability}

The concept of availability refers to repairable systems since the formula for operational availability is: \[\frac{MUT}{MUT+MDT}\] giving \emph{MUT} as the average duration of good operation after repair and \emph{MDT} the average duration of failure including the detection of the failure, the duration of intervention, the time of repair and the time to return to service. These are indeed parameters related to the fact that the equipment is repairable.

\begin{figure}[H]
\centering
\includegraphics[width=0.8\textwidth]{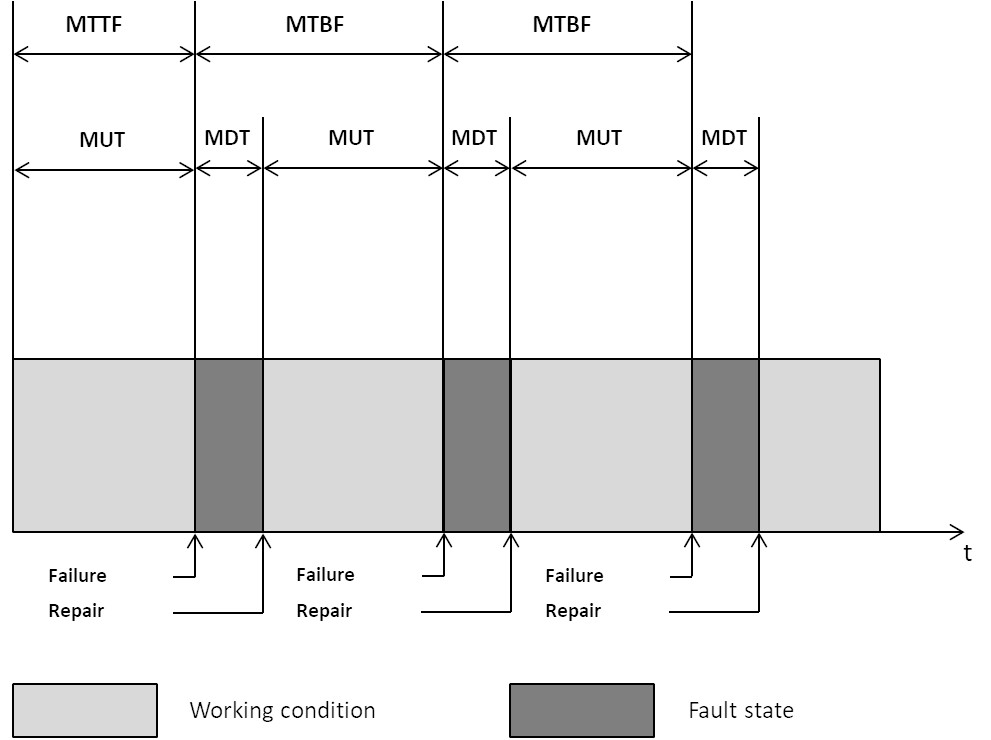}
\caption{\label{fig:MeanTime}Mean Times definition}
\end{figure}

Definitions of average times during the life of a system:

\begin{itemize}
  \item \emph{MTTF} (Mean Time To Failure) or \emph{MTTFF} (Mean Time To First Failure): average time for proper operation before the first failure.
  \item \emph{MTBF} (Mean Time Between Failure): average time between two failures of a repairable system. This definition should not be confused with the \emph{MTBF} (Mean Time Before Failure) given by the manufacturers and which corresponds to the \emph{MTTF}.
  \item \emph{MDT} (Mean Down Time): average failure time including failure detection, intervention time, repair time and return to service time.
  \item \emph{MTTR} (Mean Time To Repair): average time to repair.
  \item \emph{MUT} (Mean Up Time): average duration of good operation after repair.
\end{itemize}

To calculate the operational availability, we have the \emph{MTBF} of each component given by the manufacturers. From these \emph{MTBF} (Mean Time Before Failure) we can calculate the availability of the component using the formula: \[\frac{MTBF}{MTBF+MDT}\]

Once the availability of all the components has been calculated, we can apply the relationships from the functional block diagrams and thus obtain the availability of the node.

\section{Maintainability}

Maintainability is the ability of a system to be maintained or restarted (AFNOR). From a quantification point of view, it is the time required for the system to return to an operational state (recovery).

There are qualitative and quantitative maintainability criteria. In this paper we will focus on quantitative values.

\pagebreak

The Mean Down Time for repairable/replaceable equipment is theoretically expressed as follows: \[MDT=MTTRes+MLDT+MADT+(1-PNRS)*TAT\]

\begin{itemize}
  \item MTTRes includes the following times:
    \begin{itemize}
      \item Time for Failure notification.
      \item The diagnostic time corresponding to the average time required to carry out the diagnostic at the level of the unit to be replaced.
      \item The removal and installation time corresponding to the average time required to remove the faulty unit and replace it with a healthy unit.
      \item The verification time corresponding to the time required to validate the maintenance intervention carried out and to verify the correct operation after the maintenance intervention.
      \item The system recovery time.
    \end{itemize}
  \item \emph{MLDT} (Mean Logistic Delay Time) represents logistical times such as waiting for spare material available in buffer stock.
  \item \emph{MADT} (Mean Administrative Delay Time) represents administrative times such as waiting for qualified personnel.
  \item \emph{PNRS} represents the Probability of Non-Rupture of the Stock. A value of $99\%$ can be taken as a goal.
  \item \emph{TAT} (Turn Around Time) represents the time required to obtain a new missing spare part in the event of a buffer out of stock.
\end{itemize}

\begin{figure}[H]
\centering
\includegraphics[width=0.8\textwidth]{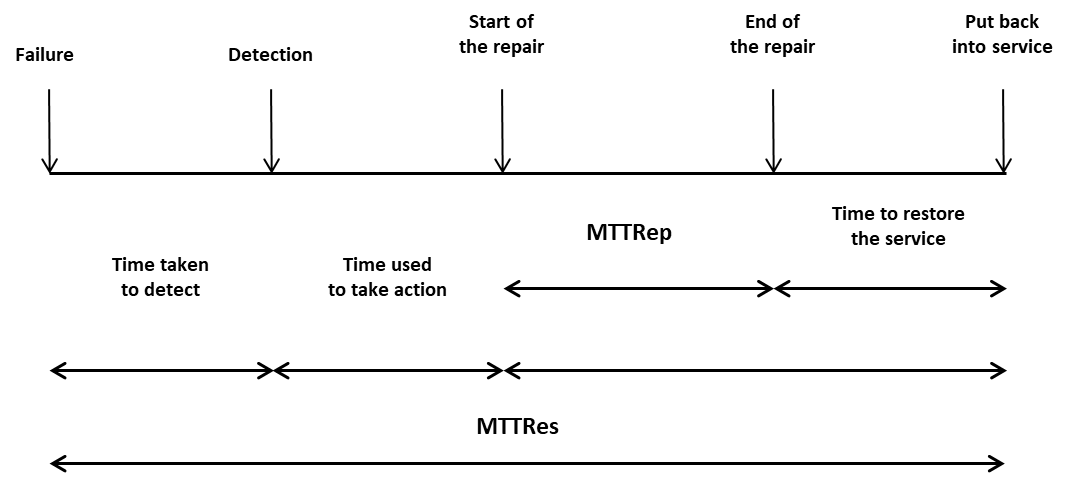}
\caption{\label{fig:Maintainability}Mean Time To Repair}
\end{figure}

\section{Reliability Block Diagrams (RBDs)}

All that remains is to reuse the Reliability Block Diagrams to calculate the availability of the system. RBDs are a method to represent a system in a simplified way and thus to decompose the system into blocks. These blocks will be series, parallel or combined components, in order to facilitate availability calculations.

\pagebreak

\subsection{Probability background}

A reminder about the probabilities will help to better understand what happens next. A probability is expressed by a number varying from \emph{1} to \emph{0}. The value \emph{1} represents the certainty that the event will happen and the value \emph{0} represents the certainty that it will not happen.

For an event $E_i : 0 \leq P(E_i) \leq 1$.

If \emph{P(A)} is the probability that event \emph{A} occurs, the probability that \emph{A} does not occur is: \[P(\bar{A}) = 1 - P(A)\]

The probability that the two events occur is the product of their respective probabilities: \[P(A \text{ and } B) = P(A) \times P(B)\]

\subsection{Generic model}

The mathematical generic model used \cite{1} is a $\frac{K}{N}$ redundancy system. A system composed of \emph{N} elements is said to have $\frac{K}{N}$ redundancy if \emph{K} elements are sufficient to carry out its mission. Redundancy is generally assumed to be active.

\begin{figure}[H]
\centering
\includegraphics[width=0.8\textwidth]{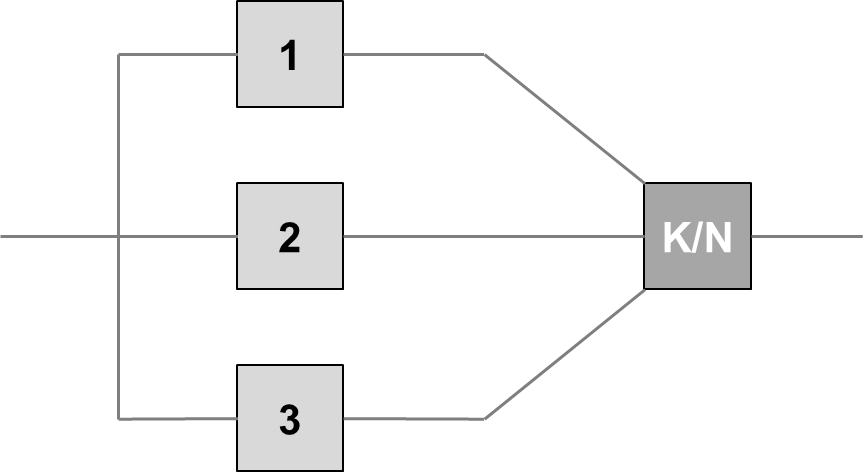}
\caption{\label{fig:GenericModel}Generic model}
\end{figure}

The availability of this system is calculated using the following formula: \[A(t) = \sum \limits_{i=K}^N C_N^i A(t)^i (1-A(t))^{N-i}\]

\pagebreak

Let's take the example of a system with three parallel elements, two of which must work for the system to work. This can be expressed in the form of probabilities:

\begin{itemize}
  \item All components must work: $A_1 \times A_2 \times A_3$
\end{itemize}

Or

\begin{itemize}
  \item At least two of the components must work: \[(1-A_1) \times A_2 \times A_3 + A_1 \times (1-A_2) \times A_3 + A_1 \times A_2 \times (1-A_3)\]
\end{itemize}

Combining all this, we get:
\begin{align}
	A(t) &= A_1 \times A_2 \times A_3 + (1-A_1) \times A_2 \times A_3 + A_1 \times (1-A_2) \times A_3 + A_1 \times A_2 \times (1-A_3) \nonumber\\
    &= A_1 \times A_2 + A_1 \times A_3 + A_2 \times A_3 - 2 \times A_1 \times A_2 \times A_3 \nonumber
\end{align}

\subsection{Series configuration}

\begin{figure}[H]
\centering
\includegraphics[width=0.8\textwidth]{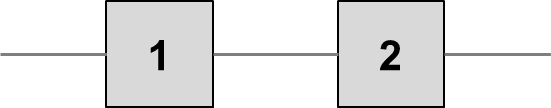}
\caption{\label{fig:SeriesSystem}Series System $(\frac{N}{N})$}
\end{figure}

A series $(\frac{N}{N})$ system is a special case in which \emph{N} elements out of \emph{N} must operate for the system to be operational.

In this case, \emph{K=N} in the generic formula: \[A(t) = \sum \limits_N^N C_N^N A(t)^N (1-A(t))^{N-N}\]

As: \[C_N^N = 1\] and \[(1-A(t))^{N-N} = 1\]

Then:
\begin{align}
	 A(t)&= A(t)^N \nonumber\\
	 &= A_1(t) \times A_2(t) \times \ldots \times A_N(t)  \nonumber\\
	 &= \prod \limits_{i=1}^N A_i(t) \nonumber
\end{align}

Indeed:
\begin{align}
    P_{(works)} &= P(A_1 \cap A_2 \cap \ldots \cap A_N)  \nonumber\\
    &= P(A_1 \text{ and } A_2 \text{ and } \ldots \text{ and } A_N)  \nonumber\\
    &= P(A_1) \times P(A_2) \times \ldots \times P(A_N) \nonumber
\end{align}

For any number $A_i(t) \in ]0, 1[$, if the values of $A_i(t)$ tend to \emph{1}, then $A_1(t) \times A_2(t) \times \ldots \times A_N(t)$ tends to \emph{0}. This series system is very unfavorable.

\subsection{Parallel configuration}

\begin{figure}[H]
\centering
\includegraphics[width=0.8\textwidth]{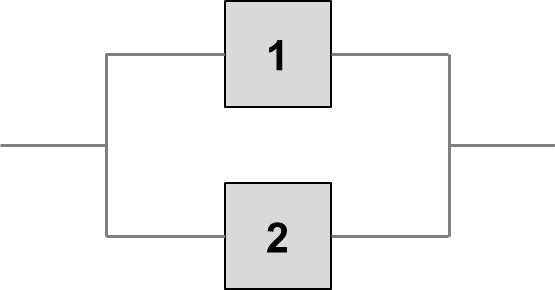}
\caption{\label{fig:ParallelSystem}Parallel System $(\frac{1}{N})$}
\end{figure}

A parallel system is a special case in which it is enough for only \emph{1} element out of \emph{N} to work for the system to be operational. In this case, \emph{K=1} in the generic formula: \[A(t) = \sum \limits_{i=1}^N C_N^i A(t)^i (1-A(t))^{N-i}\]

Which is written: \[A(t)= 1 - \prod \limits_{i=1}^N (1 - A_i(t))\]

Indeed:
\begin{align}
    P_{(works)} &= P(A_1 \cup A_2 \cup \ldots \cup A_N)  \nonumber\\
    &= 1 - P_{(does \: not \: work)}  \nonumber\\
    &= 1 - P(\bar{A}_1 \cap \bar{A}_2 \cap \ldots \cap \bar{A}_N)  \nonumber\\
    &= 1 - P(\bar{A}_1 \text{ and } \bar{A}_2 \text{ and } \ldots \text{ and } \bar{A}_N)  \nonumber\\
    &= 1 - [P(\bar{A}_1) \times P(\bar{A}_2) \times \ldots \times P(\bar{A}_N)]  \nonumber\\
    &= 1 - [(1 - P(A_1)) \times (1 - P(A_2)) \times \ldots \times (1 - P(A_N))]  \nonumber
\end{align}

For any number $A_i(t) \in ]0, 1[$, if the values of $A_i(t)$ tend to \emph{1}, $(1 - A_i(t))$ tends to \emph{0} and $(1 - A_1(t)) \times ( 1 - A_2(t)) \times \ldots \times (1 - A_N(t))$ tends to \emph{0}. Then $1 - \prod \limits_{i=1}^N (1 - A_i(t))$ tends to \emph{1}. This parallel system is very favorable.

\pagebreak

\subsection{Series-Parallel configuration}

\begin{figure}[H]
\centering
\includegraphics[width=0.8\textwidth]{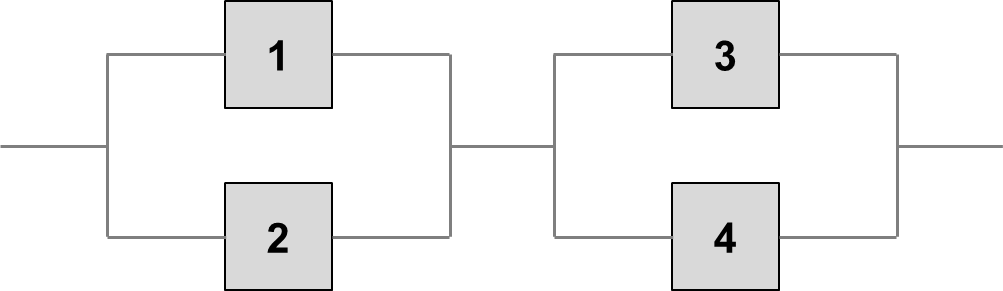}
\caption{\label{fig:SeriesParallelSystem}Series-Parallel System}
\end{figure}

\[A(t)= \prod \limits_{j=1}^P \left [ 1 - \prod \limits_{i=1}^{N_j} (1 - A_{ij}(t)) \right ]\]

with \emph{P} parallel configurations including $N_j$ elements each which are put in series and $A_{ij}(t)$ being the reliability of the $i_{th}$ element of the $j_{th}$ parallel configuration.

For any number $A_{ij}(t) \in ]0, 1[$, if the values of $A_{ij}(t)$ tend to \emph{1}, $(1 - A_{ij}(t))$ tends to \emph{0} and $(1 - A_1(t)) \times (1 - A_2(t)) \times \ldots \times (1 - A_{ij}(t))$ tends to \emph{0}. Then $\left [ 1 - \prod \limits_{i= 1}^{N_j} (1 - A_{ij}(t)) \right ]$ tends to \emph{1}. We have seen previously that the product of values which tend to \emph{1} is a result which tends to \emph{0}, so $\prod \limits_ {j=1}^P \left [ 1 - \prod \limits_{i=1}^{N_j} (1 - A_{ij}(t)) \right ]$ tends to \emph{0}. This series-parallel system is very unfavorable.

\subsection{Parallel-Series configuration}

\begin{figure}[H]
\centering
\includegraphics[width=0.8\textwidth]{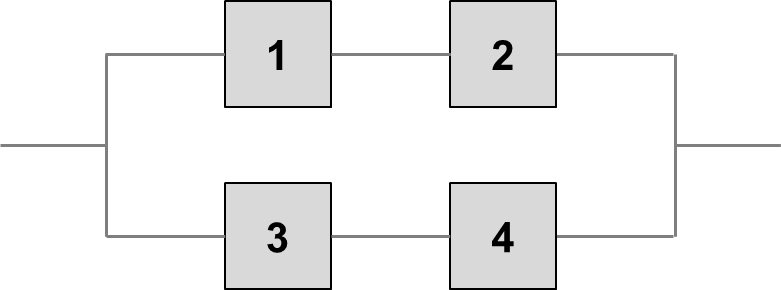}
\caption{\label{fig:ParalleSerieslSystem}Parallel-Series System}
\end{figure}

\[A(t)= 1 - \prod \limits_{i=1}^N \left [ 1 - \prod \limits_{j=1}^{P_i} A_{ij}(t) \right ]\]

with \emph{N} series configurations including $P_i$ elements each which are placed in parallel and $A_{ij}(t)$ being the reliability of the $j_{th}$ element of the $i_{th}$ series configuration.

For any number $A_{ij}(t) \in ]0, 1[$, if the values of $A_{ij}(t)$ tend to \emph{1}, $A_1(t) \times A_2(t) \times \ldots \times A_ {ij}(t)$ tends to \emph{0} and $\left [ 1 - \prod \limits_{j=1}^{P_i} A_{ij}(t) \right ]$ tends to \emph{1}. We saw previously that the product of values tending to \emph{1} is a result tending to \emph{0}, then $1 - \prod \limits_{i=1}^N \left [ 1 - \prod \limits_{j=1}^{P_i} A_{ij}( t) \right ]$ tends to \emph{1}. This parallel-series system is very favorable.

\subsection{Bridge configuration}

\begin{figure}[H]
\centering
\includegraphics[width=0.8\textwidth]{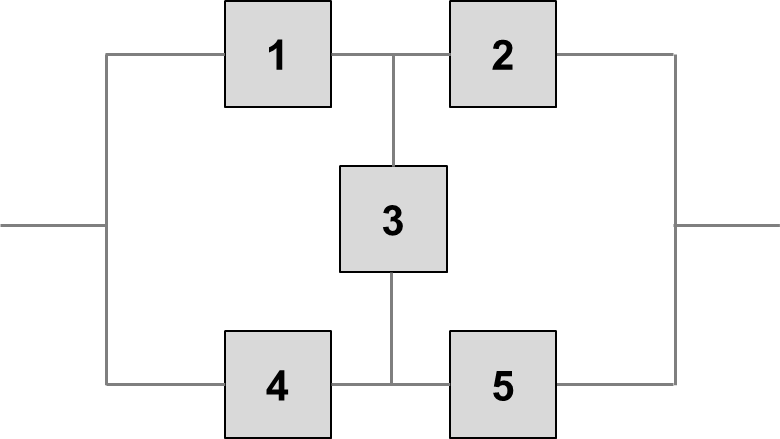}
\caption{\label{fig:BridgeSystem}Bridge System}
\end{figure}

A bridge system is not a $(\frac{K}{N})$ redundant system. This is called a system that cannot be reduced to a series-parallel combination. To compute the availability of this system, the conditional probability theorem should be used: \[A = A_3 \times A_{(3 \: works)} + (1 - A_3) \times A_{(3 \: does \: not \: work)}\]

$A(t)$ is therefore deducted from the study of the two diagrams below.

\begin{figure}[H]
\centering
\includegraphics[width=0.8\textwidth]{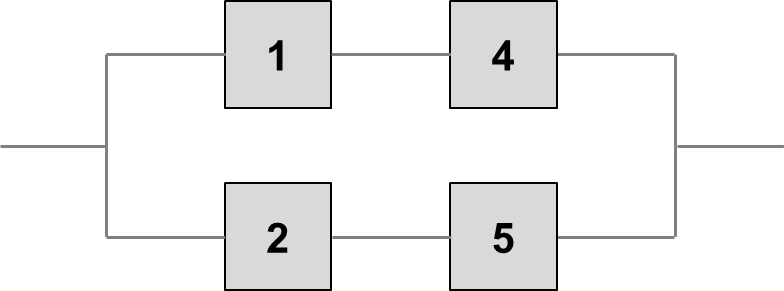}
\caption{\label{fig:ASystem}System A (3 does not work)}
\end{figure}

\begin{figure}[H]
\centering
\includegraphics[width=0.8\textwidth]{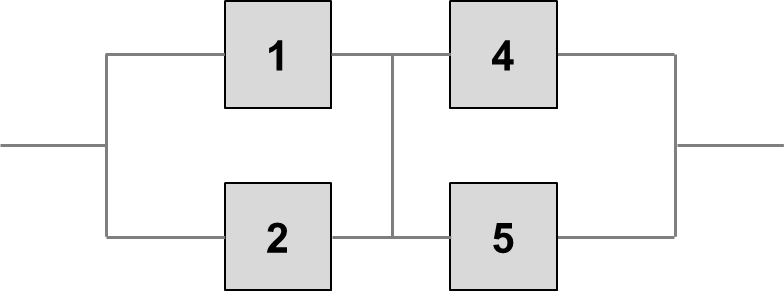}
\caption{\label{fig:BSystem}System B (3 works)}
\end{figure}

\pagebreak

From system A (parallel-series), we deduce: \[A_A(t) = 1 - (1 - A_1 \times A_4) \times (1 - A_2 \times A_5)\]

From system B (series-parallel), we deduce: \[A_B(t) = (A_1 + A_2 - A_1 \times A_2) \times (A_4 + A_5 - A_4 \times A_5)\]

Using the formula for conditional probabilities: \[A(t) = A_3 \times (A_1 + A_2 - A_1 \times A_2) \times (A_4 + A_5 - A_4 \times A_5) + (1 - A_3) \times (1 - (1 - A_1 \times A_4) \times (1 - A_2 \times A_5))\]

\bibliographystyle{alpha}
\bibliography{sample}

\end{document}